\documentclass[useAMS,usenatbib, usegraphicx]{mn2e}

\def\apj{ApJ}%
\def\apjl{ApJ}%
\def\apss{Ap\&SS}%
\def\aap{A\&A}%
\def\aaps{A\&AS}%
\def\mnras{MNRAS}%
\def\nat{Nature}%
\title[High-frequency modes in solar-like stars]{High-frequency modes in solar-like stars}
\author[C. Karoff]{C. Karoff$^{1,2}$\thanks{E-mail: karoff@phys.au.dk}\\
$^{1}$Department of Physics and Astronomy, 
	      University of Aarhus, 
              Ny Munkegade, Building 1520, 
              DK 8000 Aarhus C, Denmark\\
$^{2}$Danish AstroSeismology Centre,
	      University of Aarhus, 
              Ny Munkegade, Building 1520, 
              DK 8000 Aarhus C, Denmark\\}
\begin{document}

\date{Accepted 2007 August 1. Received 2007 July 11; in original form 2007 June 21}

\pagerange{\pageref{firstpage}--\pageref{lastpage}} \pubyear{2007}

\maketitle

\label{firstpage}

\begin{abstract}
p-mode oscillations in solar-like stars are excited by the outer convection zone in these stars and reflected close to the surface. The p-modes are trapped inside an acoustic cavity, but the modes only stay trapped up to a given frequency (known as the acoustic cut-off frequency ($\nu_{ac}$)) as modes with larger frequencies are generally not reflected at the surface. This means that modes with
   frequency larger than the acoustic cut-off frequency must be traveling waves. The high-frequency modes may provide information about the physics in the outer layers of the stars and the excitation source and are therefore highly interesting as it is the estimation of these two phenomena that causes some of the largest uncertainties when calculating stellar oscillations.

High-frequency modes have been detected in the Sun, $\beta$ Hydri and in $\alpha$ Cen A \& B by smoothing the so-called echelle diagram and the large frequency separation as a function of frequency have been estimated. The large frequency separation has been compared with a simple model of the acoustic cavity which suggests that the reflectivity of the photosphere is larger at high frequency than predicted by standard models of the solar atmosphere and that the depth of the excitation source is larger than what has been estimated by other models and might depend on the order $n$ and degree $l$ of the modes.
\end{abstract}


\section{Introduction}

Since the first observations of oscillations with frequency above the acoustic cut-off frequency in the Sun \citep{1988ssls.rept..279J, 1988ApJ...334..510L} different suggestions have been made to locate the nature of these high-frequency modes - known as  {\it"mock-modes"} \citep{1990LNP...367...87K}, {\it"high-frequency interference peaks"} \citep{1991ApJ...375L..35K}, {\it"pseudo-modes"} \citep{1995MNRAS.272..850R}, but the physic behind high-frequency modes is still not clearly understood.  

The standard way to obtain a model of the high-frequency modes is to consider a one-dimensional wave equation \citep{1990ApJ...362..256B, 1994ApJ...422L..29K}:
\begin{equation}
\frac{\rm{d}^2\psi}{\rm{dr}^2}+\left[\frac{\nu^2}{c^2}-V(r) \right]\psi=0,
\end{equation}
 where $\psi$ is the wave function and $V(r)$ is the acoustic potential. A number of studies have used this equation to make a model of the high-frequency modes. Generally the studies fall in two different categories. Either an excitation source is included on the right hand side:
 \begin{equation}
 \frac{\rm{d}^2\psi}{\rm{dr}^2}+\left[\frac{\nu^2}{c^2}-V(r) \right]\psi=\delta(r-r_0),
 \end{equation}
 and the acoustic potential is characterized by a step function \citep{1991ApJ...375L..35K, 1996ApJ...472..882A} or the excitation source is not included, but instead a more realistic model is used for the acoustic potential which includes reflection of the modes at the chromosphere-corona transition \citep{1990ApJ...362..256B, 2000A&A...361.1127D}. Using these models different studies have been able to set constraints on; the location of the excitation source \citep{1991ApJ...375L..35K}, the coronal reflection \citep{1994ApJ...422L..29K}, the solar atmosphere \citep{2000A&A...361.1127D}, the acoustic cut-off frequency \citep{2006ApJ...646.1398J}  and give possible explanations of the asymmetries of the line profiles \citep{1995MNRAS.272..850R, 1996ApJ...472..882A}. 

A third model for the high-frequency mode has been suggested by \citet{1996ApJ...456..399J}. Here is the reflection of high-frequency modes in the atmosphere caused by a horizontal magnetic field in the chromosphere with an Alfv{\'e}n speed that is a few magnitudes larger than the sound speed in the low chromosphere. Though this model is able to explain the frequency shifts observed at medium and high degree it fails at low degree where the frequency shifts predicted by the model vanish \citep{1996ApJ...456..399J} and therefore it is not investigated in this paper, as it will take decades before we can observe medium and high degree modes in solar-like stars.

Solar-like oscillations have within recent years been observed in a number of solar-like stars \citep[see][for a recent review]{2006astro.ph..9770B} and the first signs of high-frequency modes have been reported by \citet{2005ApJ...635.1281K}. Compared to observations of high-frequency modes in the Sun observations in solar-like stars have the disadvantages that the data have lower signal-to-noise ratio (S/N) and that it is only possible to observe low degree modes as it is only possible to perform disk integrated observations of the solar-like stars. Due to the low S/N it is not possible to try to fit a model to the power spectrum or the phase shifts as has been done for  Sun \citep[see e.g.][]{1991ApJ...375L..35K, 1994ApJ...428..827K, 2006ApJ...646.1398J}. Instead the frequency shifts can be used as an input for the models as they can be obtained even at low S/N. 

\citet{2005ApJ...635.1281K} were the first to observe signs of high-frequency modes in another solar-like star -- i.e. $\alpha$ Cen B. This was done by smoothing the power spectrum two times with box averages with different widths. In this way \citet{2005ApJ...635.1281K} could measure the large separation of $\alpha$ Cen B up to 7 mHz. 

High-frequency modes can prove to be very substantial for modeling oscillations in solar-like stars. The reason for this is that
high-frequency modes will be more affected by the structure in the surface layer of the star than modes in  the ordinary p-mode regime \citep{1988Natur.336..634C}. Therefore high-frequency modes have the possibility to provide valuable information of these surface layers -- e.g. sound speed and density profiles in the outer layers. This would be highly valuable to asteroseismology of solar-like stars in general as improper modeling of the surface layers is believed to cause the largest uncertainties in the frequencies of the ordinary p-modes \citep{1988Natur.336..634C}.

Following \citet{2005ApJ...635.1281K} high-frequency modes have been observed in the Sun, $\beta$ Hydri and in $\alpha$ Cen A \& B by smoothing the power spectrum, but this smoothing has been done in the echelle diagram of the power spectrum instead of in the power spectrum itself.

The paper is arranged as follows: In $\S$~2 the observations used in this paper is discussed as well as the different noise level of the observations. $\S$~3 presents a detailed analysis of the high-frequency modes in the different stars. A description of and comparison to a simple theoretical model is presented in $\S$~4 and concluding remarks are found in $\S$~5.

\section{Data}
Four different stars have been analysed in this study. The stars are the Sun, $\beta$ Hydri and $\alpha$ Cen A \& B. For all the four stars the measurements are made from disk integrated velocity observations.
\subsection{Sun}
This data set is a 805 day series of full-disk velocity observations taken by the GOLF instrument (Global Oscillations at Low Frequencies) on the \emph{Solar and Heliospheric Observatory} (\emph{SOHO}) spacecraft
\citep{2000A&A...364..799U,2005A&A...442..385G}. The data set has been calibrated as described in \citet{2005A&A...442..385G} (The data set is obtained from { \texttt http://golfwww.medoc-ias.u-psud.fr/access.html}). After removing all zero measurements the noise level in the amplitude spectrum is 0.95 mm/sec.! 

The data set consists of 3.477.600 non-zero measurements with a sampling of  20 s. This gives a Nyquist frequency of $\nu_{NY}=\frac{1}{2 \delta t} = 25.000$ $\mu$Hz and a frequency resolution of $\delta \nu=\frac{1.5}{\Delta T}=0.02$ $\mu$Hz \citep{1978Ap&SS..56..285L}.

According to \citet{1999ApJ...510L.149N} high-frequency modes are expected to have a higher S/N in photometry than in velocity, but as only velocity data are available for the three solar-like stars we have chosen also to use velocity data for the Sun. 

\subsection{$\beta$ Hydri}
$\beta$ Hydri was observed in 2005 September at the European Southern Observatory in Chile with the use of HARPS (High Accuracy Radial velocity Planet Searcher) at the La Silla 3.6 m telescope and at Siding Spring Observatory in Australia with the use of UCLES (University College London Echelle Spectrograph) at the 3.9 m Anglo-Australian Telescope (AAT) \citep{Bedding}. The data set consists of 3957 measurement and the weights have been manipulated in order to downweigh bad data points thereby increasing the S/N in the power spectrum as described by \citet{2004ApJ...600L..75B}. This has caused a noise level of 3.2 cm/sec in the amplitude spectrum at high frequencies.

The Nyquist frequency and the frequency resolution are not well defined for irregular sampled observations \citep{1999A&AS..135....1E}. We have evaluated the Nyquist frequency from a histogram of the time intervals as in \citet{1999A&AS..135....1E} and obtained a Nyquist frequency of 7200 $\mu$Hz. The frequency resolution were calculated as the FWHM of the central peak in the window function to 1.5 $\mu$Hz

Here we have only looked at frequencies below 2800 $\mu$Hz as the HARPS data contain a large artificial peak at 3070 $\mu$Hz that is due to a periodic error in the guiding system \citep{2007A&A...470..295B}. 

\subsection{$\alpha$ Cen A}
$\alpha$ Cen A was observed in 2001 May with UVES (UV-Visual Echelle Spectrograph) at the 8.2 m Unit Telescope 2 of the Very Large Telescope (VLT) and UCLES at AAT \citep{2004ApJ...600L..75B}. The weights have also been manipulated in this data set and a noise level of 1.9 cm/sec has been obtained at high frequencies in the amplitude spectrum. The data set contains 8182 measurements and we obtain a Nyquist frequency of 26.000 $\mu$Hz and a frequency resolution of 3.8 $\mu$Hz.

\subsection{$\alpha$ Cen B}
$\alpha$ Cen B was observed in 2003 May with UVES at VLT and UCLES at AAT \citep{2005ApJ...635.1281K}. By manipulating the weights a noise level of 1.3 cm/sec was obtained in the amplitude spectrum at high frequencies. \citet{2005ApJ...635.1281K} were also the first to report signs of high-frequency modes in a solar-like star other than the Sun. The data set contains 5021 measurements and we obtain a Nyquist frequency of 18.000 $\mu$Hz and a frequency resolution of 1.6 $\mu$Hz. 

\section{Data Analysis}
The power spectra of the four data sets were calculated as a weighted least-squares power spectrum \citep{1976Ap&SS..39..447L, 1995A&A...301..123F}. In order to see the high-frequency modes the four power spectra have been smoothed with a gaussian running mean, with at width equal to the large separation of the given star (values are listed in Table 1). The smoothed power spectra which have all been normalized by setting the highest peak equal to 1 are shown in Fig. 1

In order to quantify when peaks can be associated to high-frequency modes the acoustic cut-off frequency $\nu_{ac}$ has been calculated following \citet{1995A&A...293...87K},  and assuming that the derivative of the density scale height is small, then $\nu_{ac}$ will scale as $c/H$, where $c$ is the sound speed and $H$ is the density scale height expected to scale as $T/g$. In this way $\nu_{ac}$ is estimated as:
\begin{equation}
\nu_{ac}\propto \frac{M}{R^2\sqrt{T}}
\label{ac}
\end{equation}
where $M$ is the stellar mass, $R$ the radius and $T$ is the temperature. The acoustic cut-off frequency of the Sun is 5.3 mHz \citep{1990ApJ...362..256B}. The theoretical calculated acoustic cut-off frequency of the other solar-like stars as well as the mass, radius, effective temperature and large separation is shown in Table~\ref{tab1} and marked in Figs. 1, 2, 3, 4.

Just by looking at the power spectrum in Fig. 1 the high-frequency modes are visible in the Sun and to some less extend in $\alpha$ Cen B, but not in $\alpha$ Cen A and $\beta$ Hydri. 

The visibility of the high frequency modes increase significantly by averaging a number of power spectra calculated from a number of sub samples of the total time series \citep[see][]{2006ApJ...646.1398J, 1998ApJ...504L..51G}. Though the sub samples can be as short as a few days it has not been possible to preform this kind of analysis on the three solar-like stars as the entire time series is need in order to get high enough S/N. We have therefore analysed the high frequency modes in the echelle diagrams as this kind of analysis could be preformed on all four data sets.

The echelle diagrams of half the large separation are produced simply by folding the power
spectra with half the large separation as it is shown in Fig. 2. The large separations that were used were: Sun 135 $\mu$Hz \citep{2005ApJ...635.1281K}, $\beta$ Hydri 58 mHz \citep{Bedding}, $\alpha$ Cen A 106 mHz \citep{2004ApJ...600L..75B} and $\alpha$ Cen B 162 mHz \citep{2005ApJ...635.1281K}.  

The reason to produce the echelle diagram for the half large separation is that this will cause the odd and even $l$ modes to line up in the echelle diagram. 

p-modes can be found in the echelle diagram were they will line up in vertical lines as the p-modes  fulfill the asymptotic relation. In this way p-modes are clearly seen in the echelle diagram for the Sun (Fig. 2). The visibility of the p-modes is low for $\alpha$ Cen A and they are not visible in the echelle diagram for  $\alpha$ Cen B and $\beta$ Hydri (Fig. 2). The low visibility is caused by a relative low S/N for the p-modes in these stars.

In order to see the high-frequency modes the echelle diagrams have been smoothed with a Gaussian PSF. This is a technique that is well known from image manipulation -- one increases the contrast in an image by defocussing it. By smoothing the echelle diagrams the resolution gets lower, but the contrast gets higher. This means that it is not possible to see small frequency separations as e.g. rotation splitting in the smoothed echelle diagrams. Instead it is possible to see structures at low S/N as e.g. the high-frequency modes.  

The same Gaussian PSF have been used for smoothing all 4 data sets:
\begin{equation}
PSF(x,x_0,y,y_0)=e^{-\left(\frac{x-x_0}{a}\right)^2-\left(\frac{y-y_0}{b}\right)^2}
\end{equation}
where $a$ and $b$ are constants that have been set to  $a=\Delta \nu/16$ and $b=8$ echelle orders (The expression echelle order refers to one horizontal line in the echelle diagram of the half large separation).  The values of $a$ and $b$ were optimized in order to get the highest S/N for the high-frequency modes. Small changes to $a$ and $b$ did not change the behavior of the large frequency separation as a function of frequency.

In order to find peaks in the smoothed echelle diagrams the same analysis has been applied to all the four data sets. Each echelle order has been normalized by the minimum value in the echelle order. This means that the colour code in Fig. 3 gives the S/N in the given echelle order. In this way it is also possible to see structure in the echelle diagram at high frequencies and on the other hand it is possible to see if almost no structure exists -- as is the case for the Sun at high frequencies. 

The peak amplitude in each echelle order has been found by taking the centroid (centre-of-energy) in a segment of length 40 $\mu$Hz. The peaks are easily found in the ordinary p-mode regime. The peaks in higher echelle order are found by placing the middle of the segment where the peak was identified in the echelle order just below. This method is free of any
individual analysis of the data sets. The peaks found at low frequencies are not reliable, because of the smoothing of the echelle diagram. When the
power spectrum is folded with half the large separation it is
assumed that $l$=0 to 3 modes fall on top of each other and this is not the case
at low frequencies. This is clearly seen in the echelle diagram of the
Sun in Fig. 2. This means that the method outlined here cannot be
used for identifying peaks in the echelle diagram at frequencies lower
than the region of the ordinary p-modes, which is anyway not the subject
of this paper.

Some of the p-modes are seen twice in Fig. 3 as the range is large than $\Delta \nu /2$ on the horizontal axis. This means that p-modes are seen more than
once in each echelle order. As the peaks move to higher frequencies
for higher echelle orders peaks form the lower echelle order will
start appearing in the left of the images. 

The uncertainties have been estimated as:
\begin{equation}
\sigma = \sqrt{\frac{\rm{N/S}\Gamma}{4\pi}}
\end{equation}
where $\sigma$ is the uncertainty and $\Gamma$ is the FWHM of the peak. Though this formulation must be considered as empirical it is based on the discussion in \cite{1992ApJ...387..712L}. An analytical formulation of the uncertainties is not trivial as the power spectra have been folded with a PSF. 

$\Delta \nu$ for the four stars has been calculated by taking the difference between two peaks in the echelle diagram and multiplying it by two. Fig. 4 shows $\Delta \nu$ as a function of frequency for the four stars. Here it is clearly seen that the large separations
increase dramatically for frequencies just above the acoustical cut-off
frequency. One exception here might be $\beta$ Hydri where the
increase is not significantly within the uncertainties. $\beta$ Hydri is
significantly larger than the other stars therefore the peak amplitude
of the p-modes appears at lower frequency than for the other stars,
which might explain why the large separation as a function of frequency is
different for $\beta$ Hydri that for the other stars.


\section{Theory}
Below we will now compare the observed high-frequency modes to the model developed by \cite{1998MNRAS.298..464V}. This model is a simplified version of the models analysed by \citet{1991ApJ...375L..35K, 1996ApJ...472..882A} and \citet{1995MNRAS.272..850R} which means that it shows the same basis features in the power spectrum, but it allows an evaluation of which basis parameters can be extracted from the high-frequency modes. The model consists of a harmonic wave emitted from a source just below the photosphere that suffers multiple reflections at the stellar surface. The observed power spectrum of this simple model is:
 \begin{equation}
 \vert \psi \vert^2=\frac{2 \pm 2\rm{cos}(4\pi\nu T_s)}{1+R(\nu)^2-2R(\nu)\rm{cos}(4\pi\nu T)}
 \end{equation} 
 where $T$ is the acoustic depth of the acoustic cavity, $T_s$ is acoustic distance between the lower reflection point and the source ($\Delta T=T-T_s$ equals the acoustic distance from the source to the upper reflection) and $R(\nu)$ is the reflection coefficient as a function of frequency. If the source is a monopole then the sign in the numerator will be a plus sign and minus for a dipole source. 

The reflection coefficient depends on the acoustic potential which can be approximated with a parabolic profile \citep[see][and references herein]{1998MNRAS.298..464V}. This gives a reflection coefficient as a function of frequency as:
\begin{equation}
R(\nu)=\frac{1}{1+\rm{exp}\frac{\nu^2-\nu^2_{ac}}{\nu^2_0}}
\end{equation}
where $\nu_{ac}$ is the frequency of the maximum value in the parabolic potential profile (roughly equal to the acustic cut-off frequency) and $\nu_0$ determines the width of the parabolic barrier.

The free parameters in the model are: $\Delta T$, and $\nu_0$ -- i.e. the position of the excitation source and the reflection coefficient in the outer layers of the star.  The two parameters $T$ and $\nu_{ac}$ are not considered as free parameters as observations of ordinary p-modes can set relatively tight constrains on these two parameters. $T$ determines the large separation of the ordinary p-modes and $\nu_{ac}$ can be estimated either by the scaling low relation given by \citet{1995A&A...293...87K} or by a bivariate analysis (coherence and phase shift) as it is done by \citet{2006ApJ...646.1398J}. Models have therefore only been made with different $\Delta T$, and $\nu_0$. Comparison from the models to the stars other than the Sun can easily be made just by scaling the acustic cut-off frequency and the mean large separation.

The large separation can easily be obtained in these noise free power spectra simply by identifying the highest point in each peak. These large separations are shown as a function of frequency for eight sets of $\Delta T, \nu_0$ in Fig. 5. The other values used in the models are the same as used by \citet{1998MNRAS.298..464V} -- i.e. $T$ = 1000 s, $\nu_{ac} = 5$ mHz and a dipole source.

Fig. 5 shows that the complexity of the structure of the large separation increases as $\Delta T$ decreases and as $\nu_0$ increases. The physics behind these causal relations is explained by \citet{1998MNRAS.298..464V}. When the excitations source is moved outwards (as $\Delta T$ decrease) the frequency of a global trapped mode needs to be higher for the outermost node to be near the source. Therefore as $\Delta T$ is decreased the bumps in Fig. 5 will move to higher frequencies. However, at higher frequencies energy leakage due to reduced reflection become important, therefore the amplitudes of the bumps are lowered as the bumps move to higher frequencies. If, on the other hand the reflectivity is increased at higher frequencies (as $\nu_0$ increase) the amplitudes of the bumps are increased as energy leakage is reduced. This can been seen in the large separation plots in Fig. 5.

Simulations have been made with both a monopole and a dipole source with the same conclusion as found by \citet{1998MNRAS.298..464V} -- i.e. that the bumps in Fig. 5 will move to higher frequencies when going from a monopole to a dipole source. If one compares Fig. 4 to Fig. 5 it is seen that the observations are in favor of a dipole source, as the bumps generally appear at higher frequencies in the observations than in the simulations.

Looking at Fig. 4 bumps are found in the Sun (at 6000 and 7500 $\mu$Hz), $\alpha$ Cen A (at 4100 and 4900 $\mu$Hz) and $\alpha$ Cen B (at 5400 and 6800 $\mu$Hz).  For the Sun and $\alpha$ Cen A the two bumps appear at frequencies higher than $\nu_{ac}$ for $\alpha$ Cen B the second bump appear at the same frequency as $\nu_{ac}$. Comparing this with the models in Fig. 5 suggests that $\nu_0$ should have a value of 4000 $\mu$Hz or higher in the Sun and $\alpha$ Cen A. Setting $\nu_0$ = 4000 $\mu$Hz (and $\nu_{ac}$ = 5000 $\mu$Hz) results in a reflection of 0.42 at 5500 $\mu$Hz and 0.12 at 7500 $\mu$Hz. This contradicts the results by \citet{1998MNRAS.298..464V} who propose a upper limit of 0.02 for the reflectivity at 7500 $\mu$Hz. For $\alpha$ Cen  B a $\nu_0$ of 3000  $\mu$Hz is found to agree best with the observations as no clear bump apears for frequency higher than $\nu_{ac}$, but only some small curvature at 9000 $\mu$Hz. The uncertainty is estimated to $\pm$1000 $\mu$Hz and is shown together with the estimated values of $\Delta T$ and $\nu_0$ in Table 2. The estimation is based on the ability of model to reproduce the observation. The uncertainties are therefore the same for the Sun and the other stars though the S/N is different in the observations. This also means that improving the model could lower the uncertainties significantly.   

In $\beta$ Hydri a small bump is seen at 1900 $\mu$Hz, but it does not appear to be statistically significant.

As two bumps are only seen for a $\Delta T$ higher than 50 s we estimate the acoustic depth of the excitation source to 50 $\pm$ 10 s for the Sun. By using the sound speed in the outer layers of the Sun from Model S  \citep{1996Sci...272.1286C} this can be converted to a source depth of 460 $\pm$ 100 km (below the radius where $T$ = $T_{\rm{eff}}$) which is significantly larger than the 140 $\pm$ 60 km that is found by \citet{1994ApJ...428..827K}.  A possible reason for the discrepancy in the estimation of $\nu_0$ and $\Delta T$ could be that only low degree modes are analysed here, whereas \citet{1994ApJ...428..827K}  and \citet{1998MNRAS.298..464V} have analysed high degree modes. 

\citet{2000MNRAS.314...75C} discuss a number of different estimates of the depth of the excitation source mainly obtained from the asymmetry in the p-mode profile. Depending on the model used the depth is found to be between 75 to 1500 km (the later value is found when using the first derivatives of $G_{\Psi}$). Evidence is seen that the depth of the excitation source varies with frequency \citep{2000MNRAS.314...75C}, but there is no clear evidence that the source depth depends on the degree though the smallest values are found in analysis of  medium and high degree modes \citep{1994ApJ...428..827K, 1999ApJ...510L.149N}. 
 
The Sun shows shape structures in the large separation for frequencies higher than 7200 $\mu$Hz (Fig. 4). These shape structures are caused by the bump in the echelle diagram at 8000 $\mu$Hz (Fig. 3). By comparing to the models in Fig. 5 this bump could look like an artifact, but it has not been possible to remove it by reanalyzing the data with adjusted parameters. Though the Sun would follow the models predictions much better if the bump were removed and the bump appears in the frequency range with low S/N in the power spectrum. The same is the case for $\alpha$ Cen A for frequencies higher than 5000 $\mu$Hz. Here the line in the echelle diagram could also follow the contour at x=40 $\mu$Hz instead of the contour at 55 $\mu$Hz. This would also make $\alpha$ Cen A follow the model prediction much better, but again this could not be accomplished by reanalyzing the data with adjusted parameters.


\section{Discussion}
High-frequency modes have been detected in the Sun, $\beta$ Hydri and $\alpha$ Cen A \& B. By using a simple model of the high-frequency modes that is able to reproduce the main structure in the large separation it is possible to parameterize the model of the high-frequency modes. Using the model we find that the reflection is higher at high frequency (0.42 at 5500 $\mu$Hz and 0.12 at 7500 $\mu$Hz) than what has been found by other studies and that the excitation source is placed deeper in the Sun (460 $\pm$ 100 km) than what have been found by other studies. As most of the other studies have analysed oscillations at lower order $n$ and higher degree $l$ this indicates that the excitation of modes with different ($n$,$l$) does not take place at the same depth.

Analysis of high-frequency modes in solar-like stars has shown to be able to provide two extra parameters in addition to the frequencies of the ordinary p-modes to be used in computation of stellar models. The two extra parameters are the depth of the excitation source and the reflectivity of the stellar atmosphere. In this paper a simplified model has been used for the high-frequency modes where the reflectivity was parameterized with $\nu_0$. Of course future studies should try to compare the high-frequency modes with models using reflectivity profiles calculated from more sophisticated model of the stellar atmosphere and with models using realistic profiles for the excitation source instead of a $\delta$-function.

It has been proven in this paper that high-frequency modes can be used in computation of oscillations for solar-like stars. Great advances in our understanding of the effect of the outer layers of the stars on the oscillations can therefore be expected with the successful launch of COROT on the 27th of December 2006 \citep{2002sshp.conf...17B} and the launch of Kepler in 2008 \citep{2003SPIE.4854..129B}. These two mission will hopefully provide us with observations of high-frequency modes in a large number of solar-like stars. Here the high-frequency modes will benefit from being in a frequency range not affected by instrument noise and being observed with photometry (and not with radial velocities at is the case in this paper). According to \citet{1999ApJ...510L.149N} observing in photometry is expected to be an advance as the high-frequency modes are expected to have a higher S/N in photometry than in velocity (noise meaning here stellar noise and not instrument noise).

\section*{Acknowledgements}
I would like to thank J. Christensen-Dalsgaard, H. Kjeldsen and D.O. Gough for many useful comments on this study. I also acknowledge support from Instrument Center for Danish Astrophysics.

\begin{figure*}
\begin{center}
          \includegraphics[width=\columnwidth]{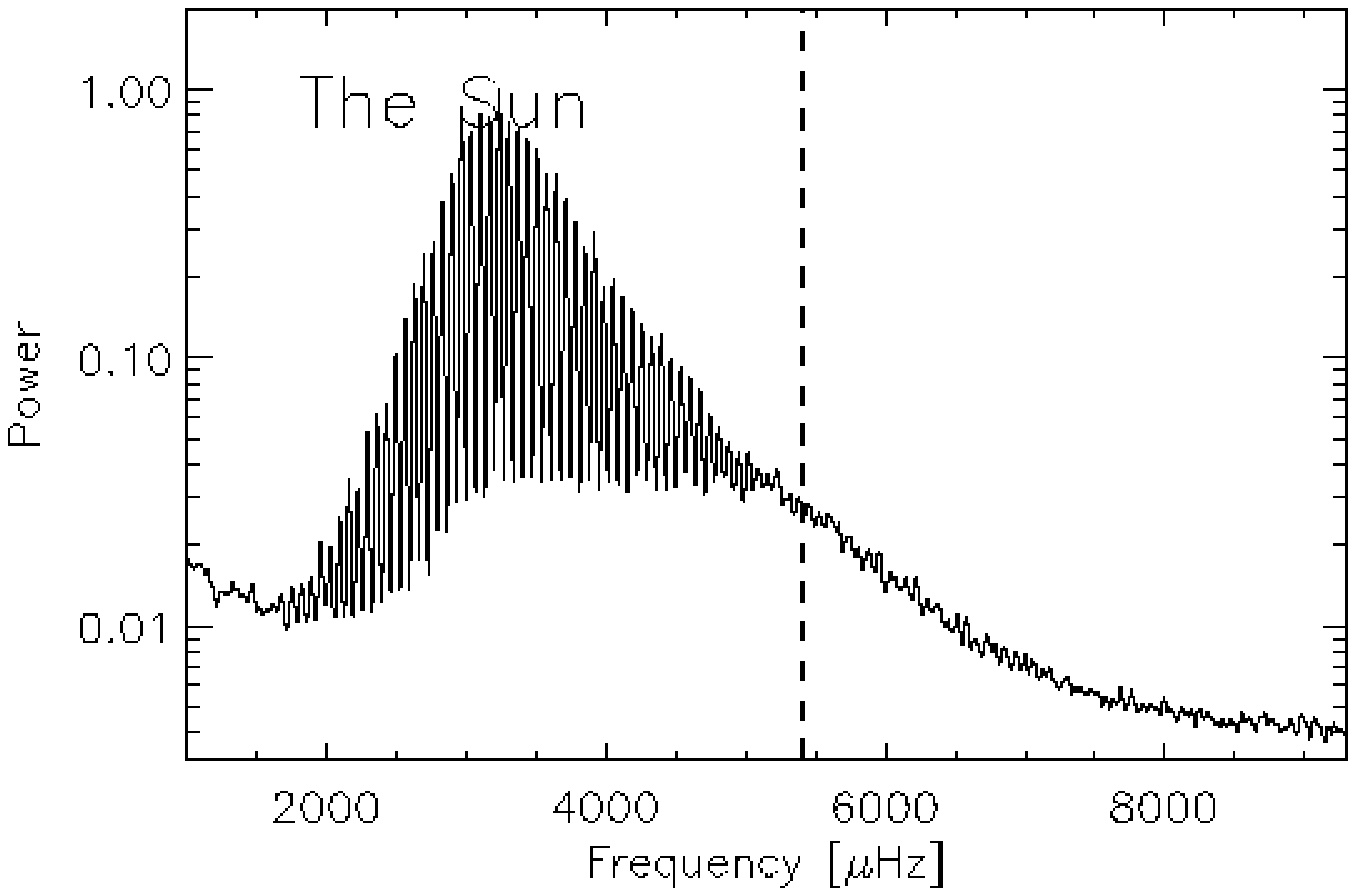}
	 \includegraphics[width=\columnwidth]{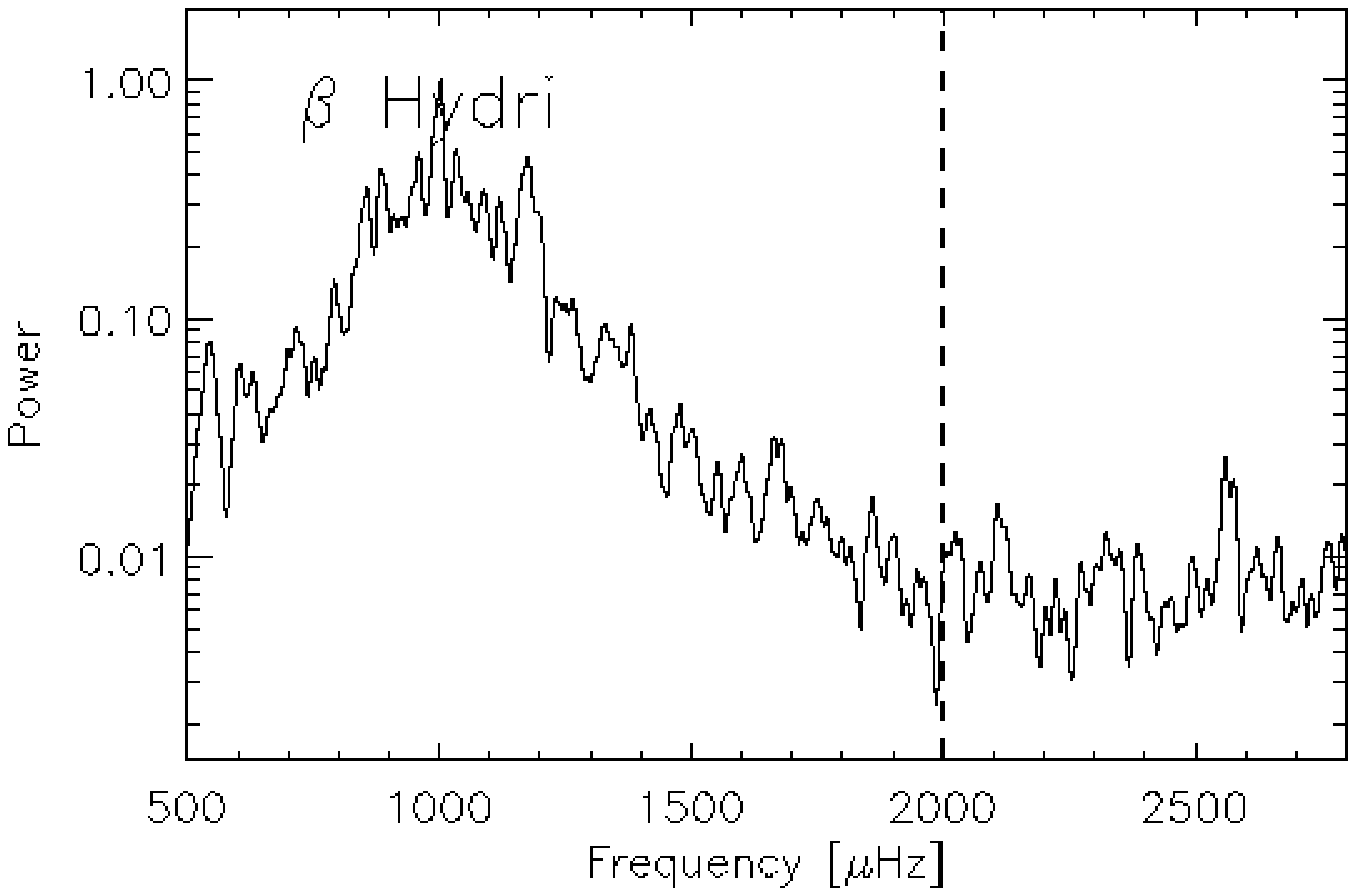}
	 \includegraphics[width=\columnwidth]{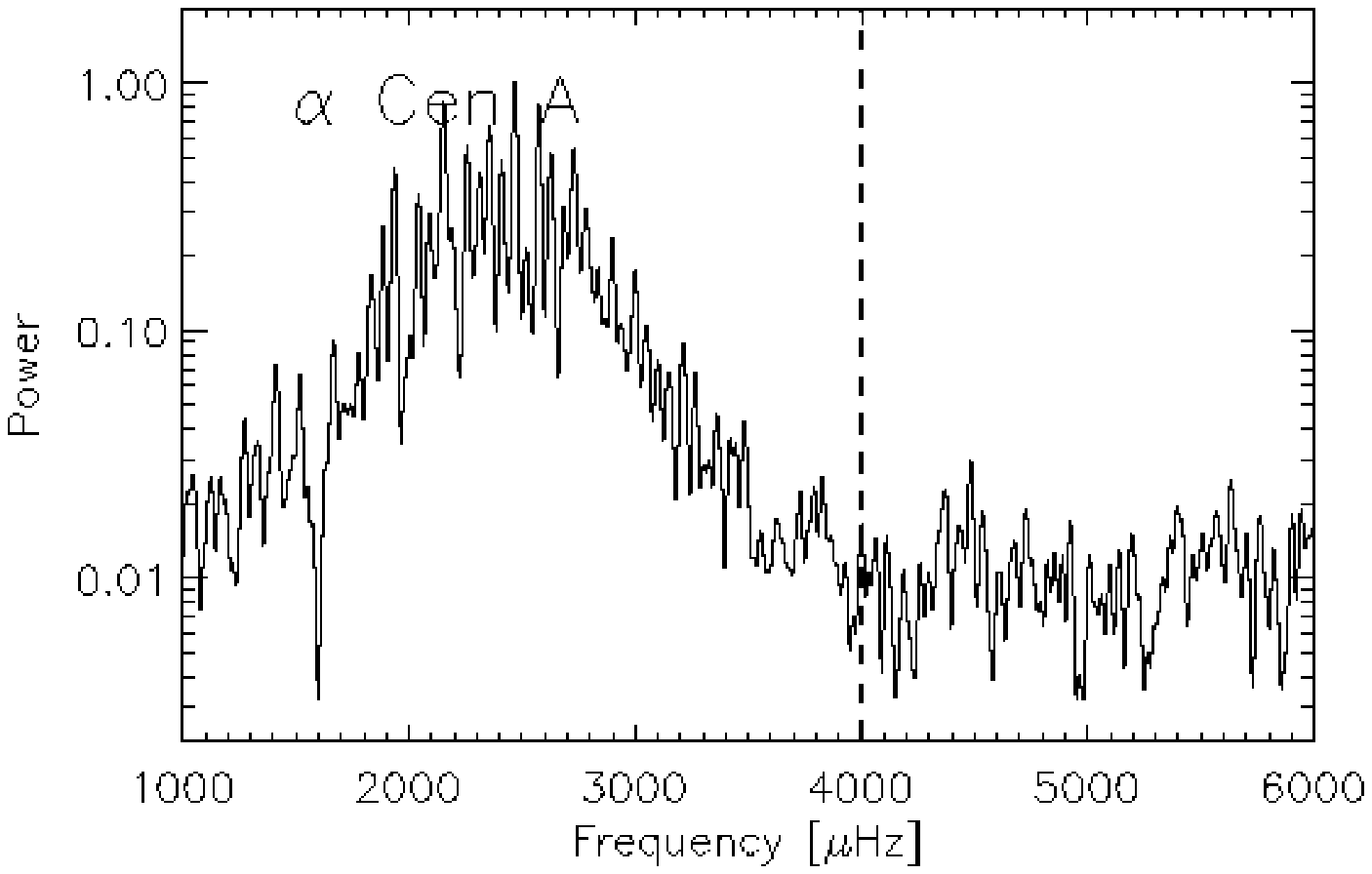}
	 \includegraphics[width=\columnwidth]{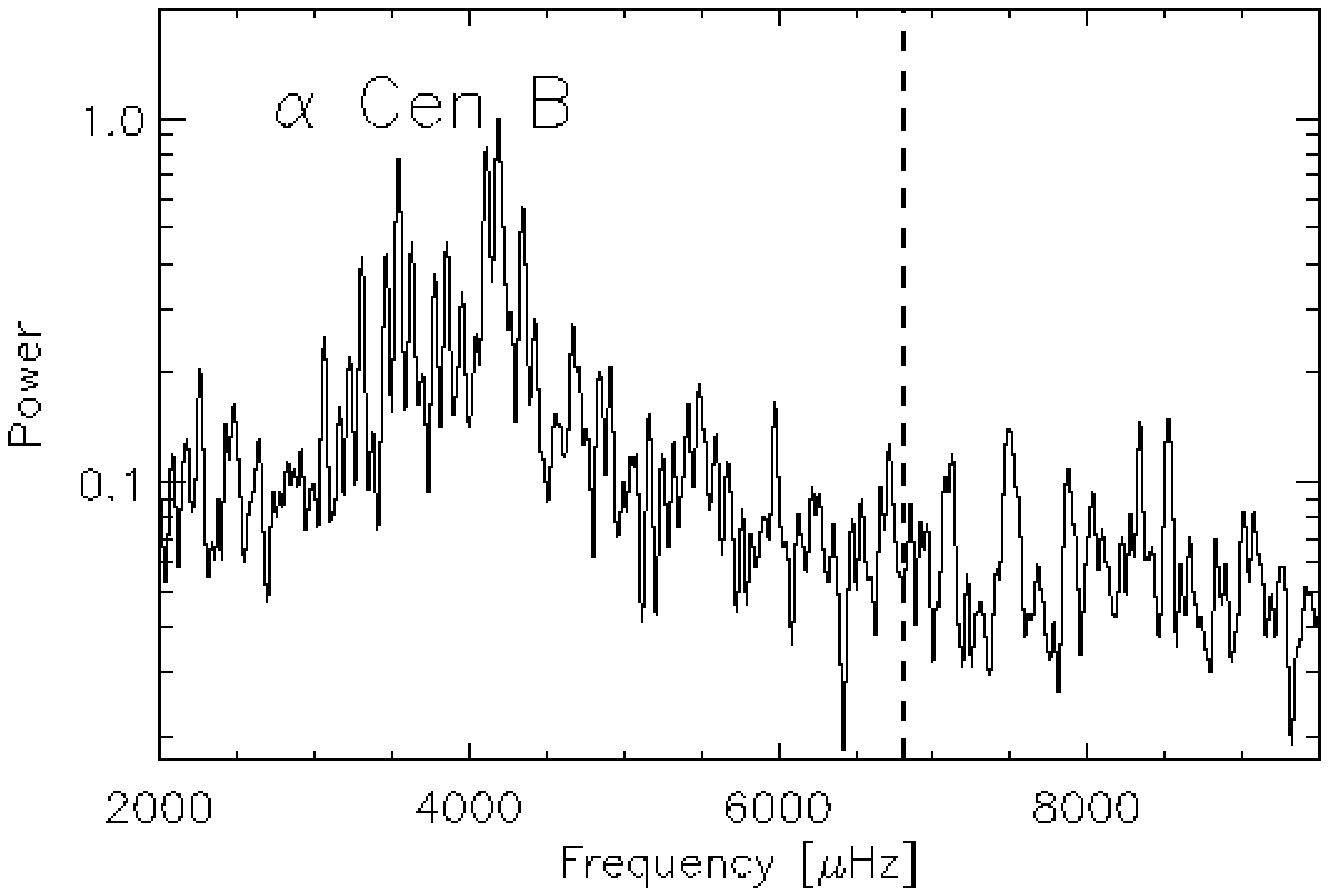}
\end{center}
\label{fig2}
\caption{Smoothed power spectrum for the Sun, $\beta$ Hydri, $\alpha$ Cen A and $\alpha$ Cen B. All the power spectras have been smoothed with a gaussian running mean with at width equal to the large separation of the given stars (values are listed in Table 1) and normalized to 1 for the highest peak. The dotted lines shown the theoretical calculated acustic cut-off frequency.}
\end{figure*}

\begin{figure*}
\begin{center}
          \includegraphics[width=\columnwidth]{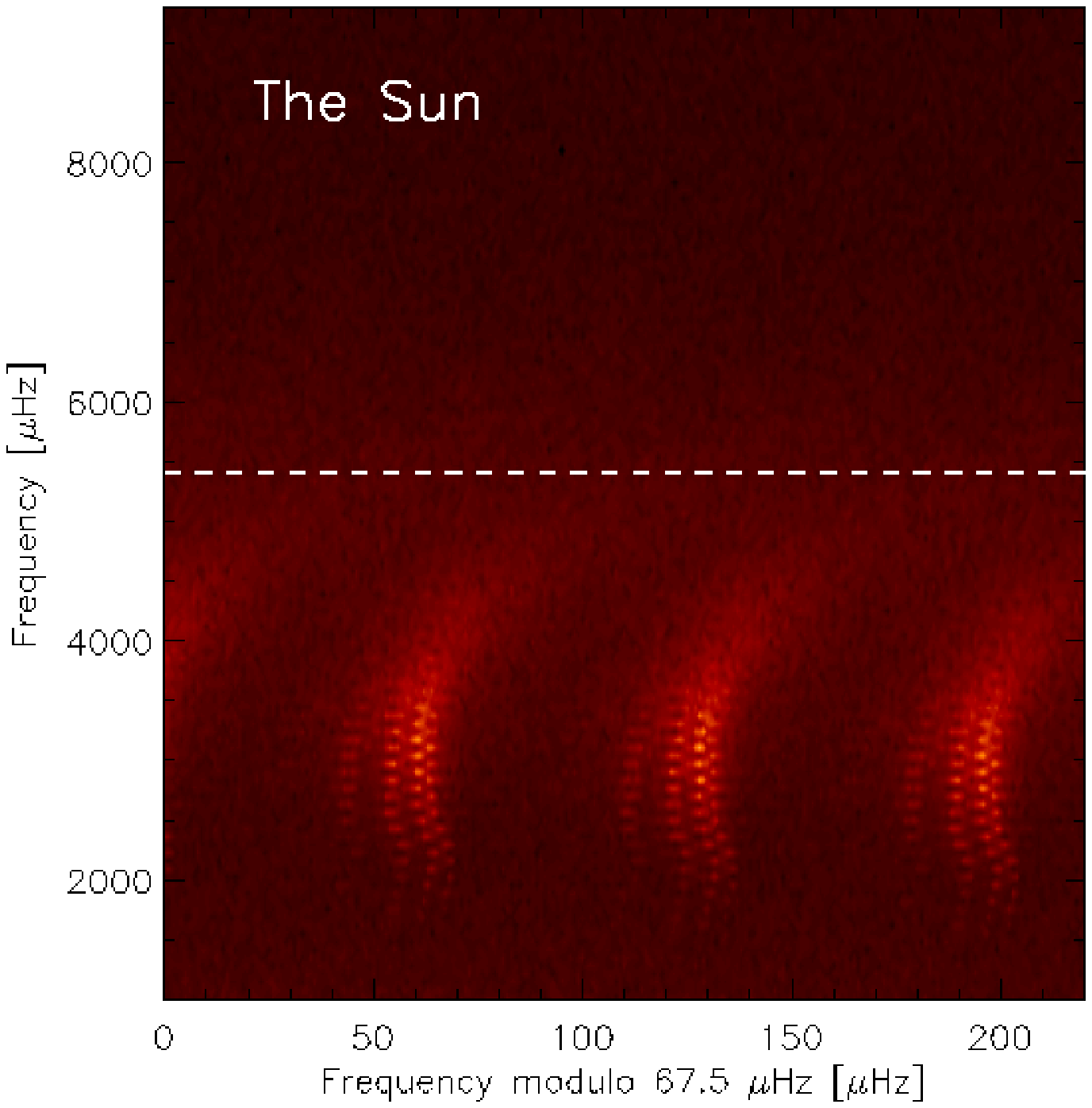}
	 \includegraphics[width=\columnwidth]{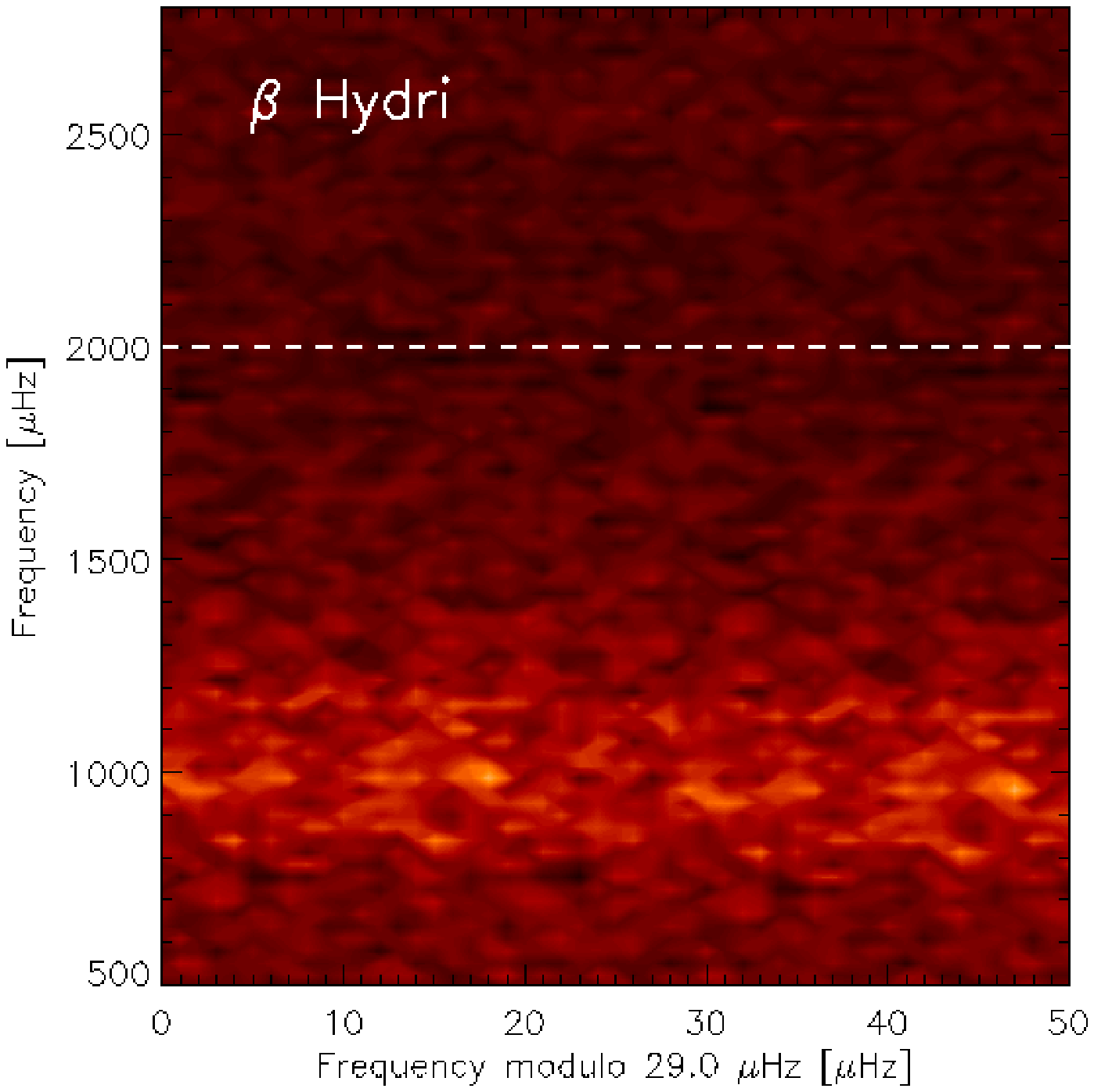}
	 \includegraphics[width=\columnwidth]{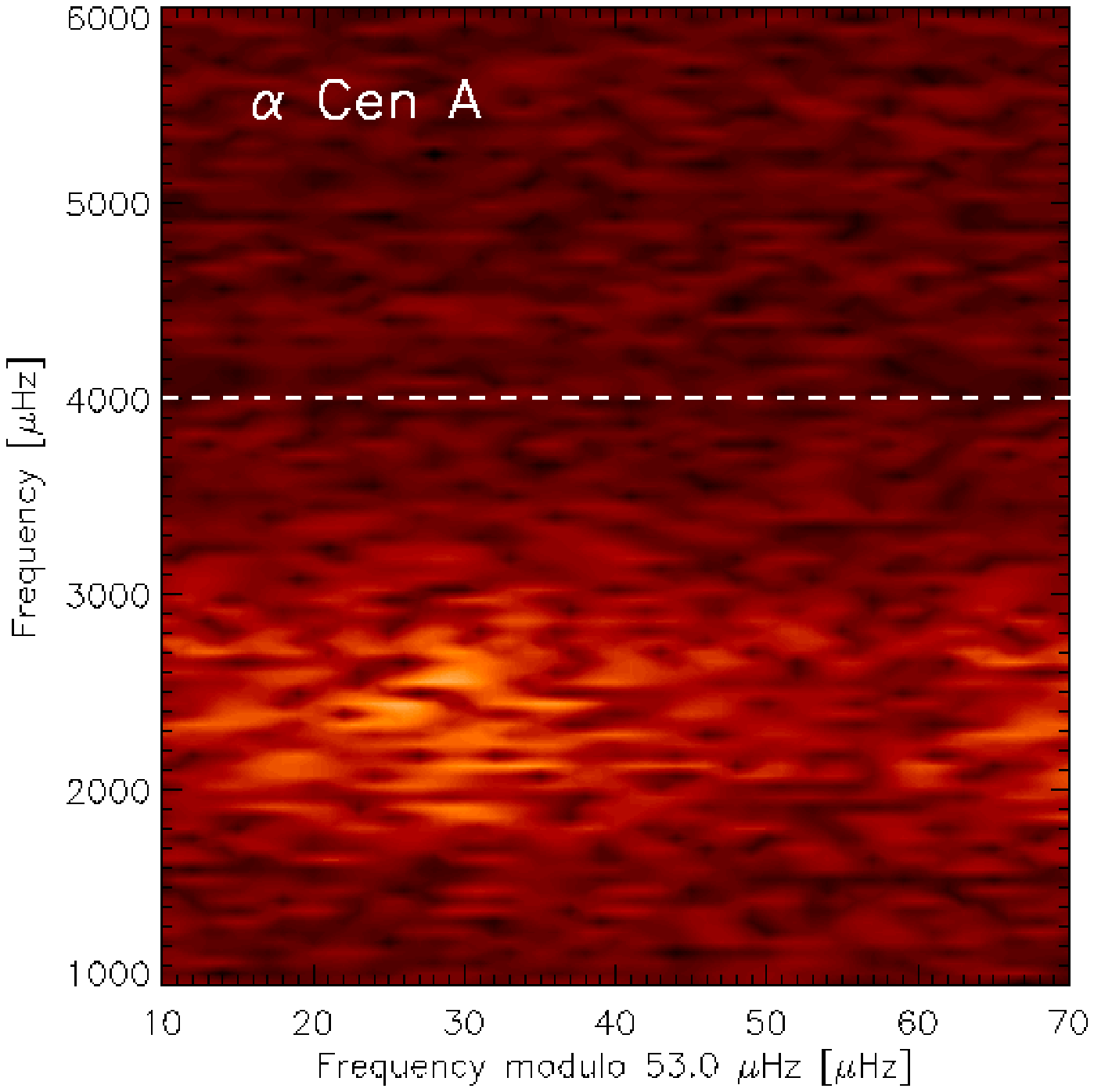}
	 \includegraphics[width=\columnwidth]{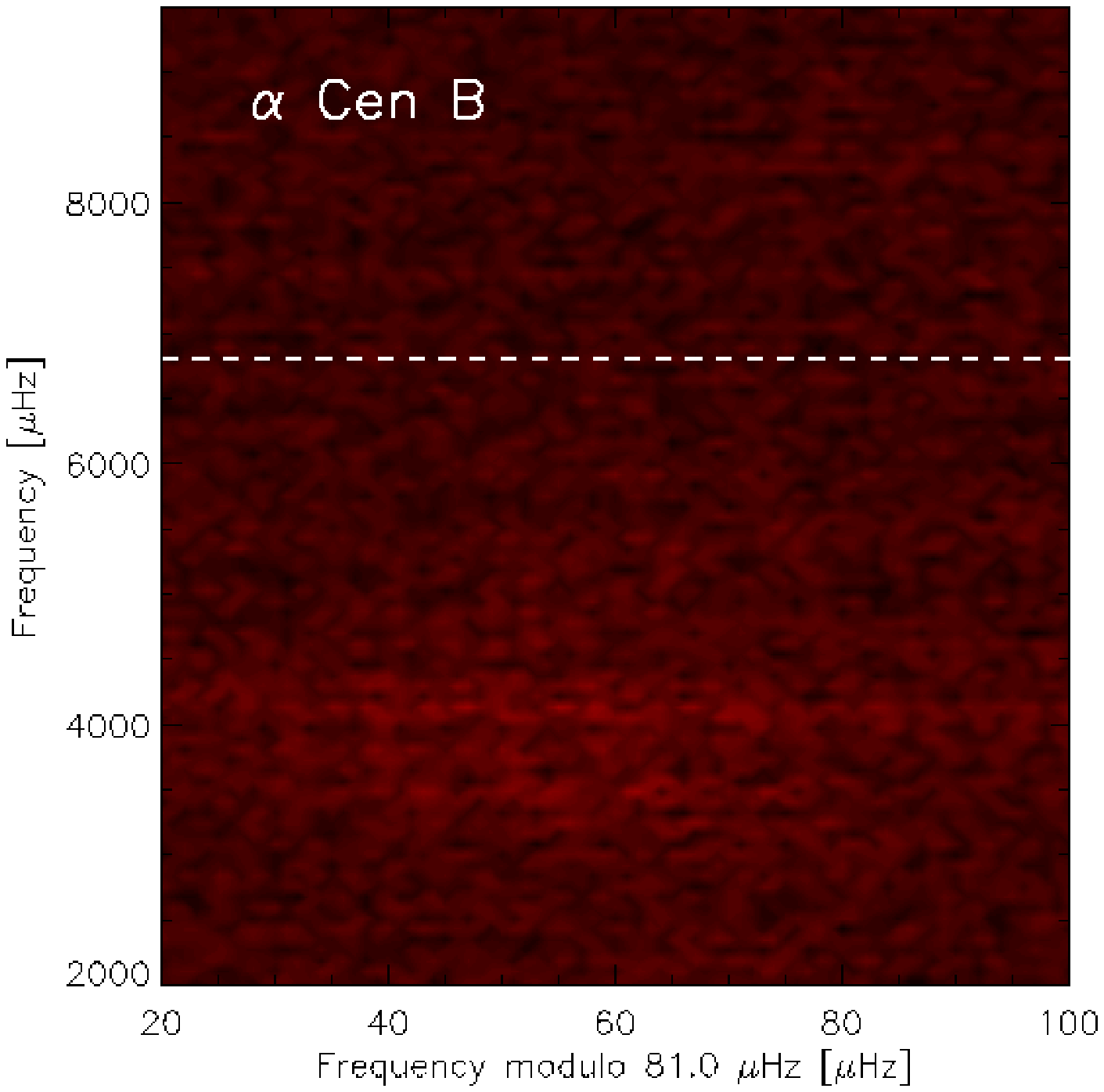}
\end{center}
\label{fig3}
\caption{Echelle diagram for the Sun, $\beta$ Hydri, $\alpha$ Cen A and $\alpha$ Cen B. The echelle diagrams are made by folding the power spectra with half the large separation. The dotted lines shown the theoretical calculated acoustic cut-off frequency.}
\end{figure*}

\begin{figure*}
\begin{center}
          \includegraphics[width=\columnwidth]{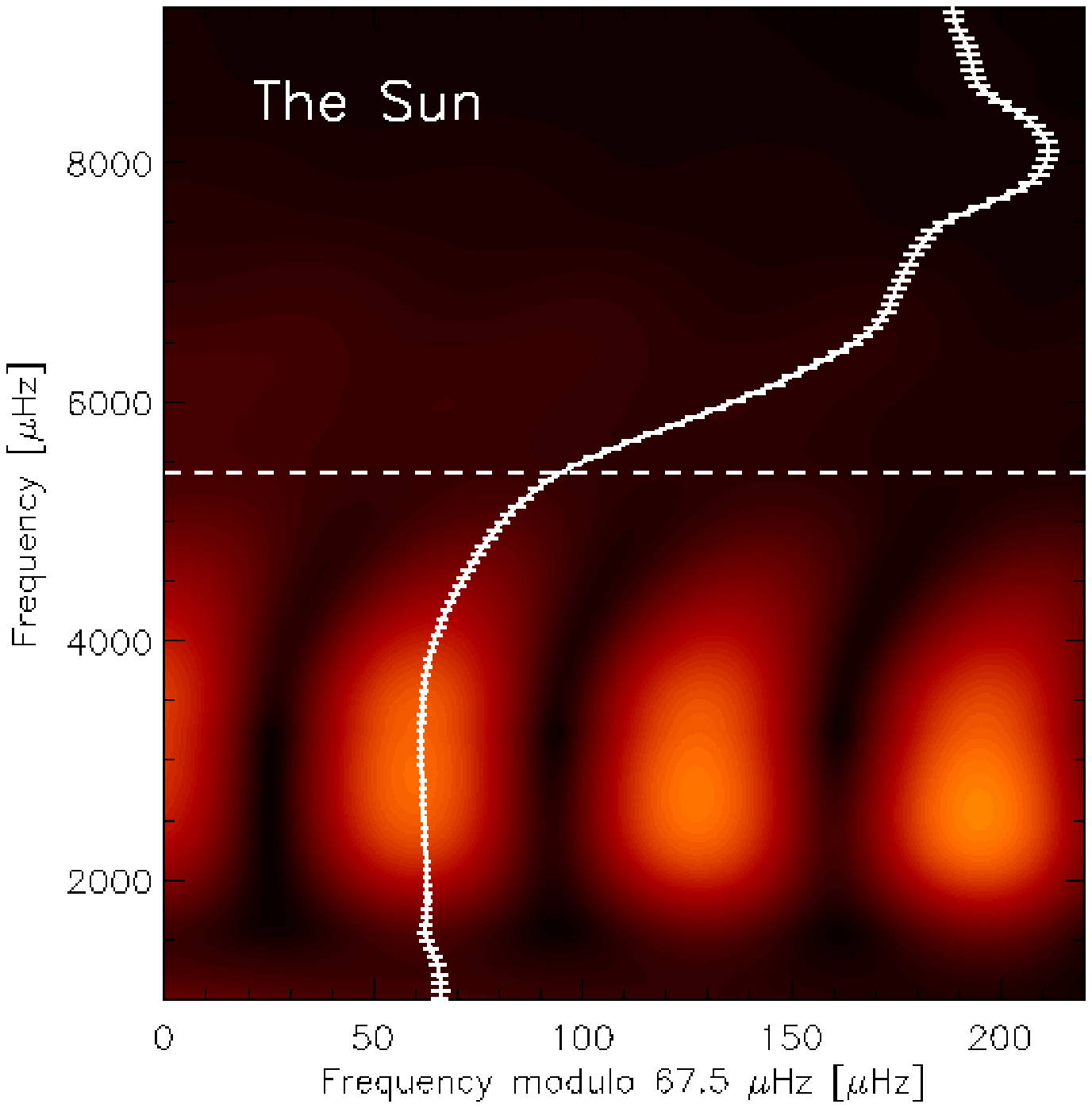}
	 \includegraphics[width=\columnwidth]{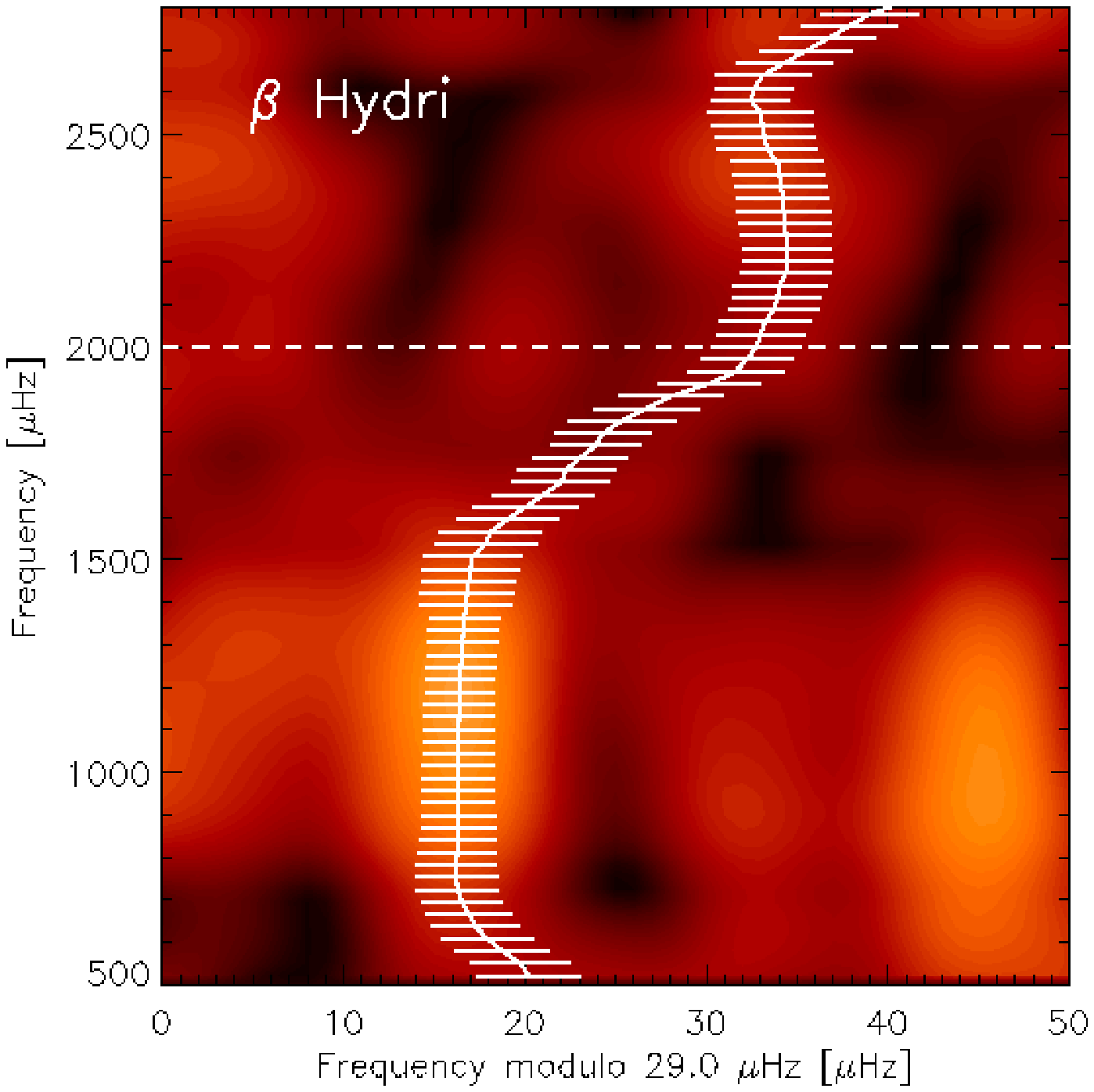}
	 \includegraphics[width=\columnwidth]{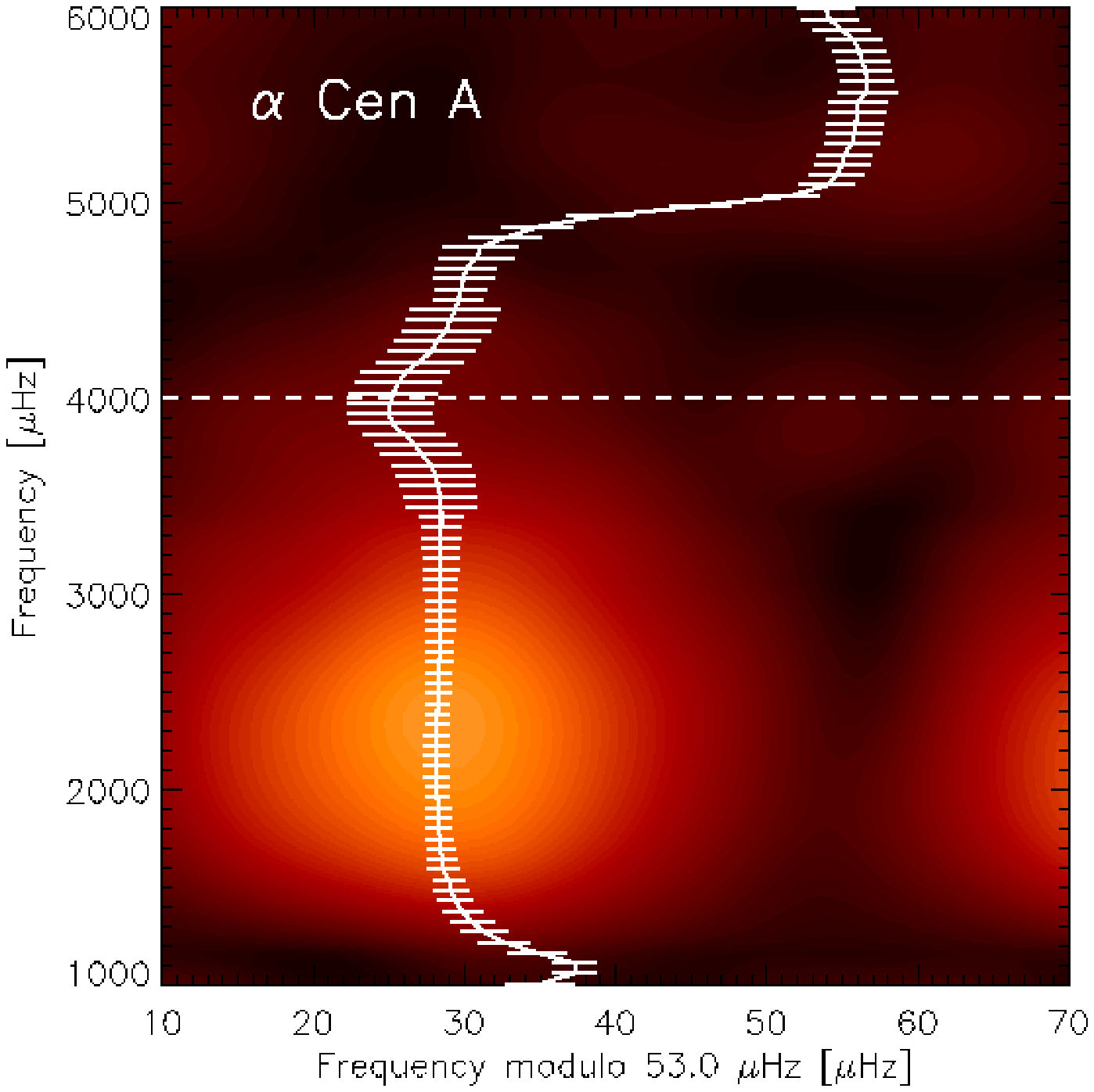}
	 \includegraphics[width=\columnwidth]{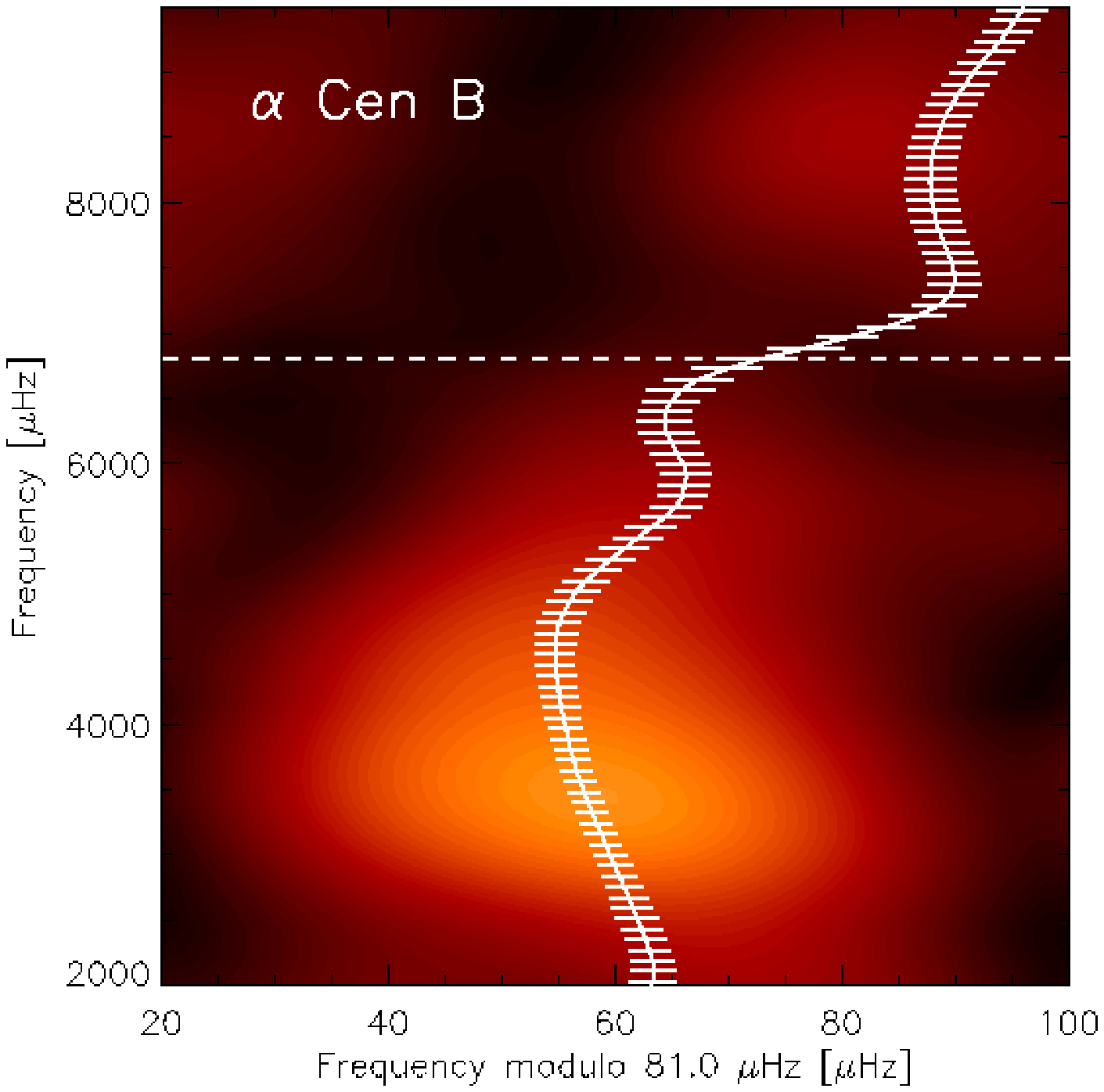}
\end{center}
\label{fig4}
\caption{
PSF folded echelle diagram for the Sun, $\beta$ Hydri, $\alpha$ Cen A and $\alpha$ Cen B. The solid line marks the obtained trend in the echelle diagram which is used for calculating the large separation. The uncertainties are calculated as described in the text. The dotted lines shown the theoretical calculated acoustic cut-off frequency.}
\end{figure*}

\begin{figure*}
\begin{center}
          \includegraphics[width=\columnwidth]{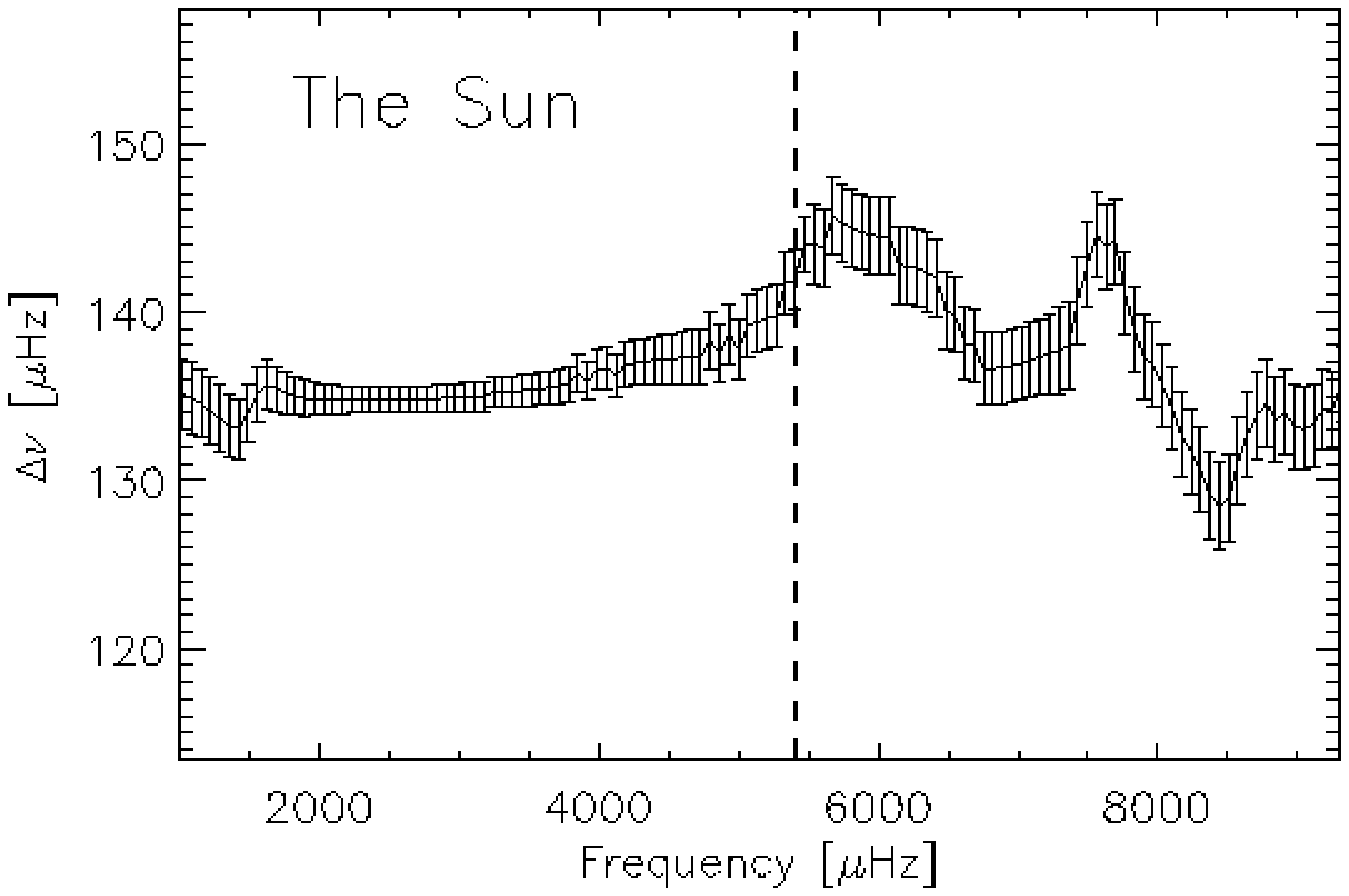}
	 \includegraphics[width=\columnwidth]{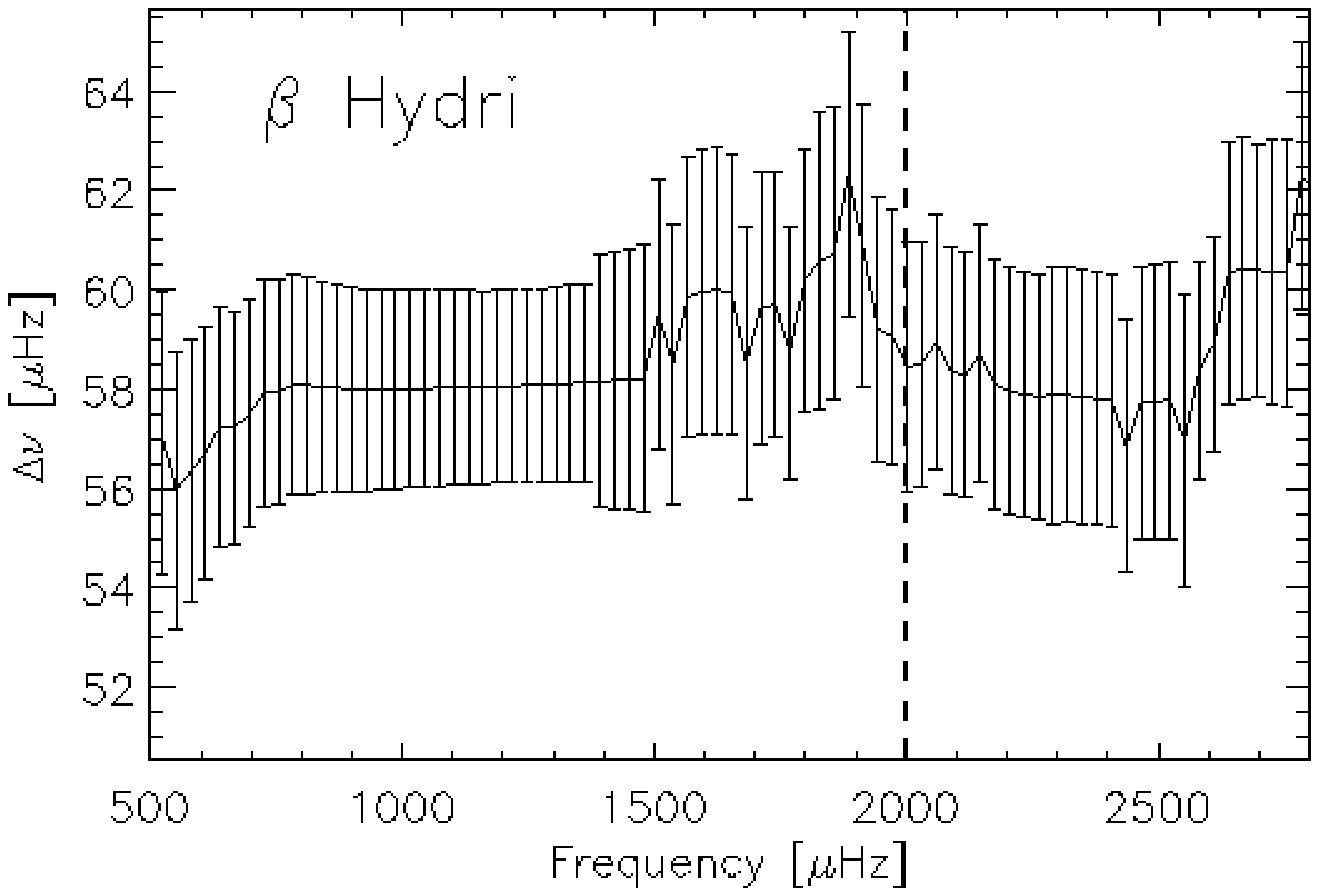}
	 \includegraphics[width=\columnwidth]{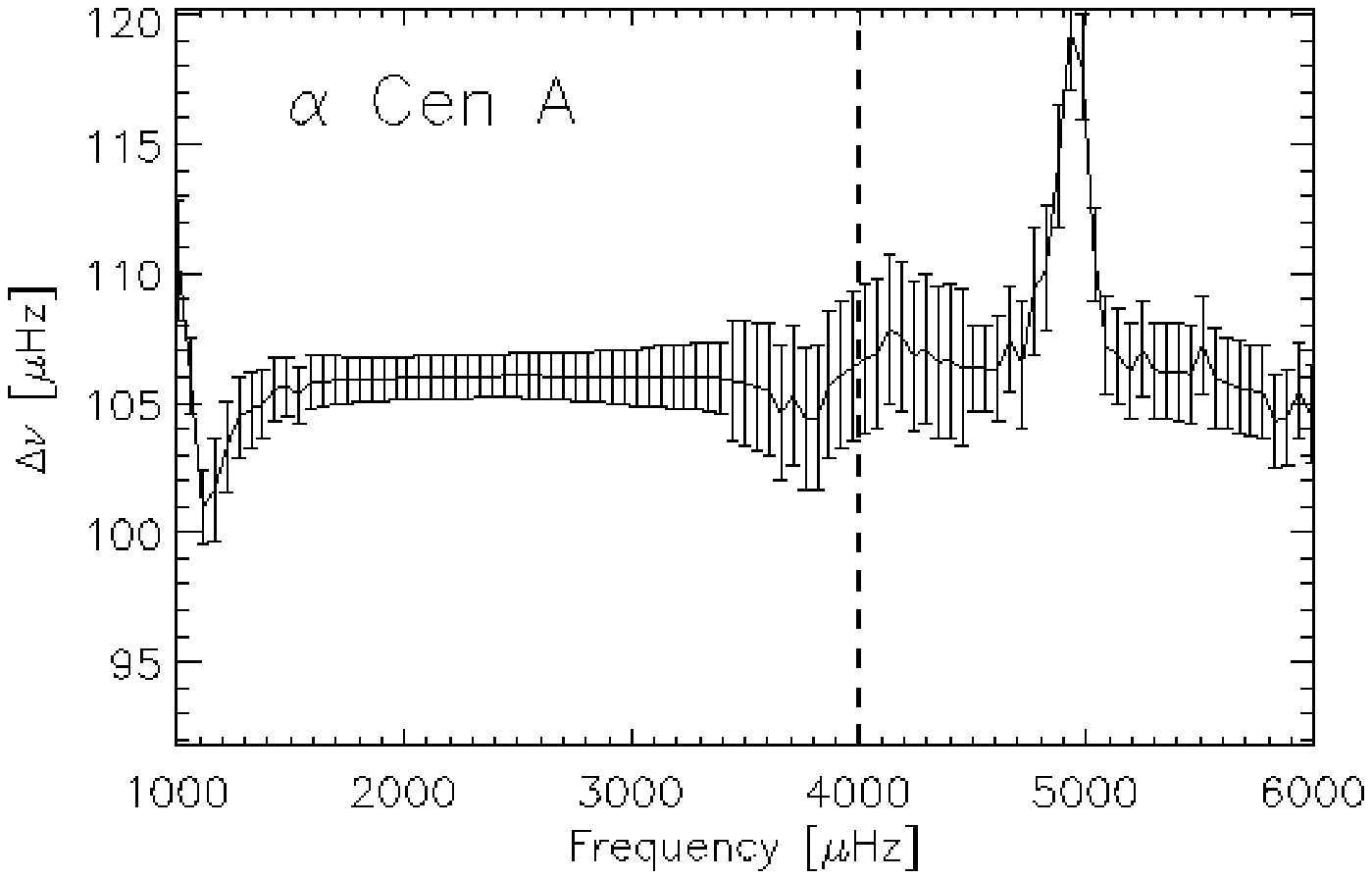}
	 \includegraphics[width=\columnwidth]{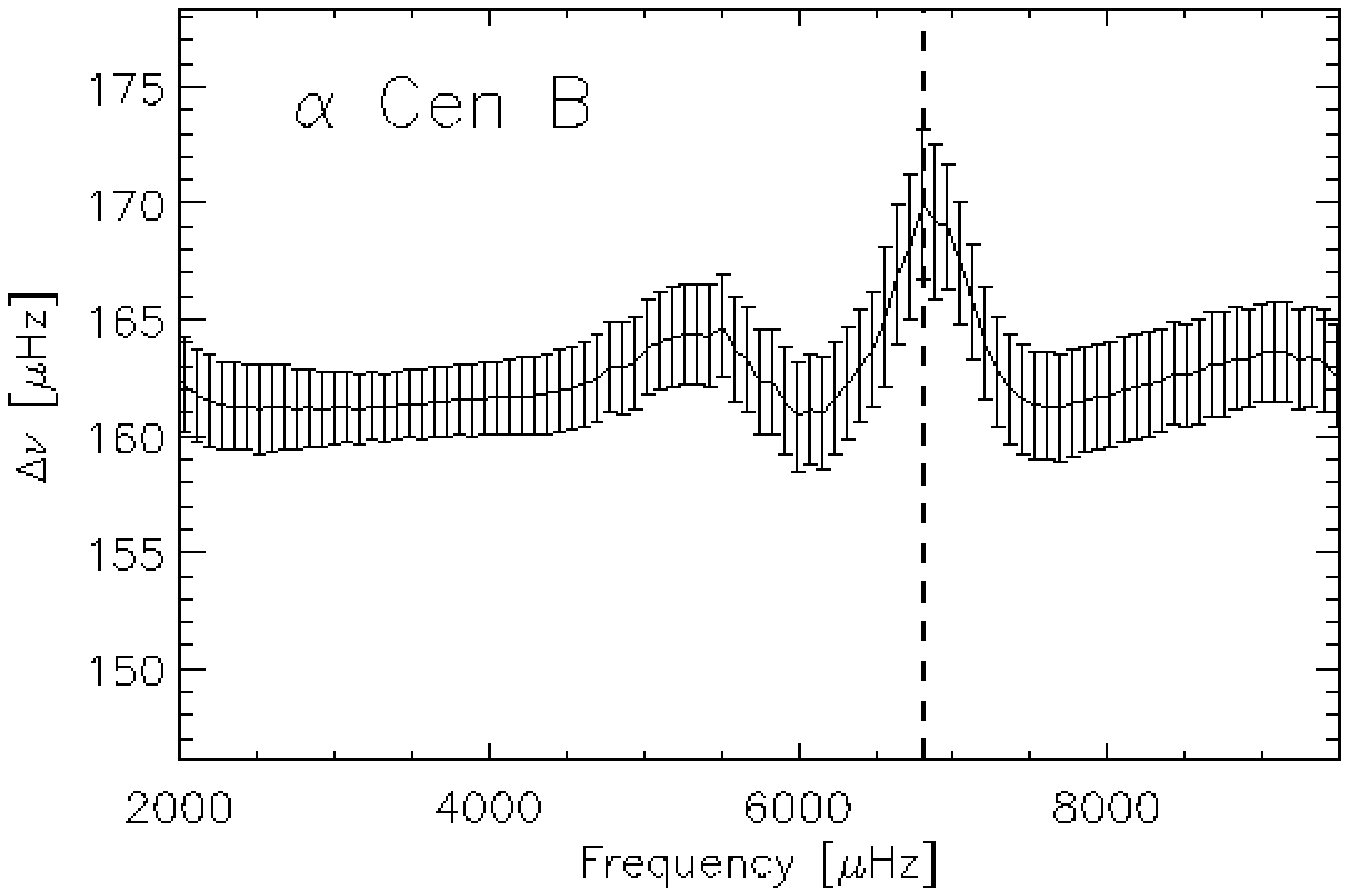}
\end{center}
\label{fig5}
\caption{The large separation as a function of frequency for the Sun, $\beta$ Hydri, $\alpha$ Cen A and $\alpha$ Cen B. The dotted lines shown the theoretical calculated acoustic cut-off frequency.}
\end{figure*}

 \begin{figure*}
\begin{center}
          \includegraphics[width=\columnwidth]{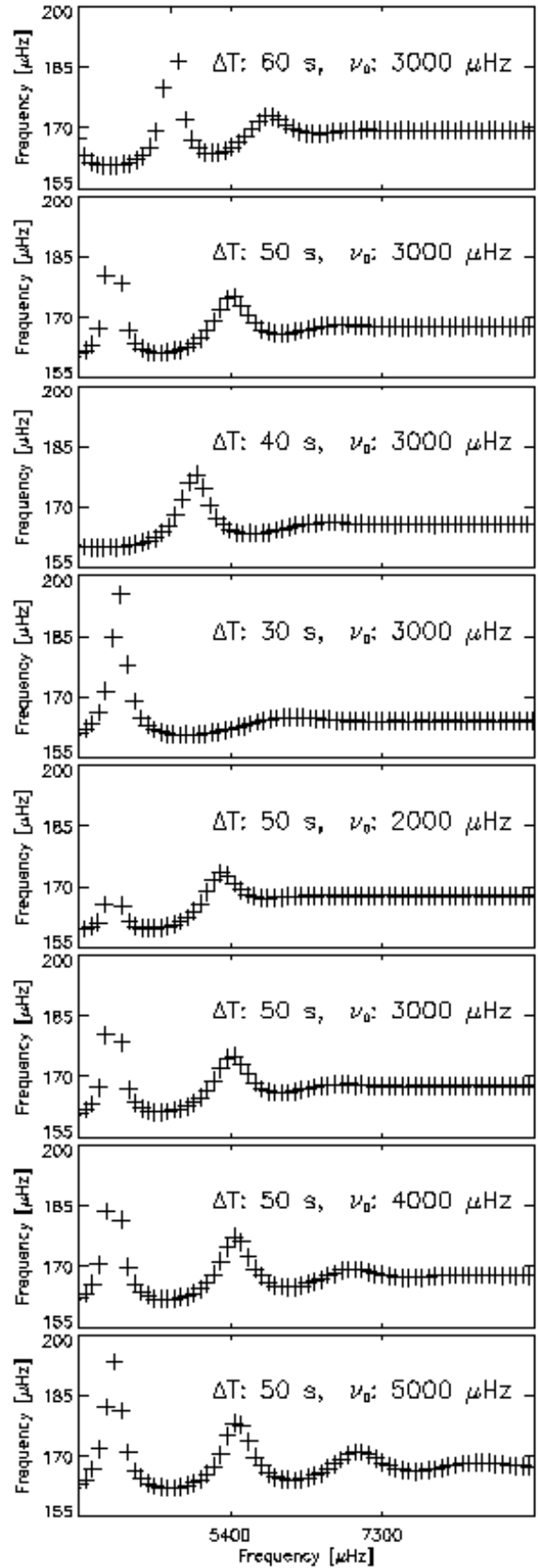}
\end{center}
\label{fig6}
\caption{Eight different models of the large separation as a function of frequency. $\Delta T$ and $\nu_0$ are typed at the top of each plot.}
\end{figure*}

\begin{table}
\caption{Stellar parameters. Ref 1: \citet{1990ApJ...362..256B} , Ref 2: \citet{2003Ap&SS.284..229D}, Ref 3: \citet{2005A&A...441..615M}}
\centering
\begin{tabular}{lcccccc}
\hline \hline
 Star & M/M$_{\odot}$ & R/R$_{\odot}$ & T$_{\rm{eff}}$ & $\nu _{ac}$ &
 $\Delta \nu$ & Ref.\\
\hline
Sun & 1.00 & 1.00 & 5777 K & 5.3 mHz & 135 $\mu$Hz & 1\\
$\beta$ Hydri & 1.14 & 2.00 & 5860 K & 2.0 mHz & 58 $\mu$Hz & 2\\
$\alpha$ Cen A & 1.11 & 1.22 & 5810 K & 4.0 mHz & 107 $\mu$Hz & 3\\
$\alpha$ Cen B & 0.93 & 0.86 & 5260 K & 6.8 mHz & 162 $\mu$Hz & 3\\
\hline
\end{tabular}
\label{tab1}
\end{table}

\begin{table}
\caption{Estimated depth of the excitation source and width of the reflection}
\centering
\begin{tabular}{lcc}
\hline \hline
 Star & $\Delta T$ & $\nu_0$ \\
\hline
Sun & 50 $\pm$ 10 s & 4000 $\pm$ 1000 $\mu$Hz\\
$\alpha$ Cen A & 50 $\pm$ 10 s & 4000 $\pm$ 1000 $\mu$Hz\\
$\alpha$ Cen B & 50 $\pm$ 10 s & 3000 $\pm$ 1000 $\mu$Hz\\
\hline
\end{tabular}
\label{tab2}
\end{table}
\bsp

\label{lastpage}

\end{document}